\documentclass[preprint,12pt]{elsarticle}

\usepackage{amssymb}

\usepackage{graphicx}
\usepackage{booktabs}
\usepackage[a4paper, margin=1in]{geometry}
\usepackage[unicode]{hyperref}
\usepackage{multirow}
\usepackage{amsmath}

\journal{arXiv}

\begin{document}

\begin{frontmatter}


\title{Diff-CAPTCHA: An Image-based CAPTCHA with Security Enhanced by Denoising Diffusion Model\tnoteref{label1}}


\title{}

\author[a]{Ran Jiang}
\author[a]{Sanfeng Zhang\corref{cor}}
\ead{sfzhang@seu.edu.cn}
\author[b]{Linfeng Liu}
\author[c]{Yanbing Peng}
\cortext[cor]{Corresponding author}
\affiliation[a]{organization={School of Cyber Science and Engineering},
            addressline={Southeast University},
            city={Nanjing},
            postcode={210018},
            country={China}}

\affiliation[b]{organization={School of Computer Science and Technology},
            addressline={Nanjing University of Posts and Telecommunications},
            city={Nanjing},
            postcode={210023},
            country={China}}


\affiliation[c]{organization={Fiberhome starrysky com\&develop CO. of Nanjing},
            city={Nanjing},
            postcode={210019}, 
            country={China}}

\begin{abstract}
To enhance the security of text CAPTCHAs, various methods have been employed, such as adding the interference lines on the text, randomly distorting the characters, and overlapping multiple characters. These methods partly increase the difficulty of automated segmentation and recognition attacks. However, facing the rapid development of the end-to-end breaking algorithms, their security has been greatly weakened. The diffusion model is a novel image generation model that can generate the text images with deep fusion of characters and background images. In this paper, an image-click CAPTCHA scheme called Diff-CAPTCHA is proposed based on denoising diffusion models. The background image and characters of the CAPTCHA are treated as a whole to guide the generation process of a diffusion model, thus weakening the character features available for machine learning, enhancing the diversity of character features in the CAPTCHA, and increasing the difficulty of breaking algorithms. To evaluate the security of Diff-CAPTCHA, this paper develops several attack methods, including end-to-end attacks based on Faster R-CNN and two-stage attacks, and Diff-CAPTCHA is compared with three baseline schemes, including commercial CAPTCHA scheme and security-enhanced CAPTCHA scheme based on style transfer. The experimental results show that diffusion models can effectively enhance CAPTCHA security while maintaining good usability in human testing.

\end{abstract}

\begin{graphicalabstract}
\end{graphicalabstract}



\end{frontmatter}


\section{Introduction}
CAPTCHA(Completely Automated Public Turing Test to tell Computers and Humans Apart) \cite{von2003captcha} is a Turing test used to distinguish between computers and humans, which is widely used to defend against automated attacks and abuse of online services. It serves as an important security mechanism in the Internet. The fundamental principle of CAPTCHA is to pose an AI challenge that requires users to provide the correct response to proceed with the services. Otherwise, the further access will be denied. A well-designed CAPTCHA scheme should be easily solved by humans while presenting enough difficulties to computer programs. Existing CAPTCHA schemes can be categorized into text-based schemes, image-based schemes, audio-based schemes, and other types\cite{xu2020survey}. Among them, the text-based CAPTCHAs are the most widely utilized by many websites, such as Wikipedia and Microsoft\cite{wang2023improving,bursztein2011text}.

However, in recent years, there have been some breakthrough advancements in deep learning-based recognitions for images and text, allowing attackers to use image classification models such as ResNet\cite{he2016deep} and EfficientNet\cite{tan2019efficientnet}, as well as object detection models such as Faster R-CNN\cite{ren2015faster}, to identify the CAPTCHA images. Attacks on the text-based CAPTCHAs often involve obtaining enough labeled dataset and training a classification model to automatically recognize the CAPTCHAs. For the pure text-based CAPTCHAs, the success rate of end-to-end attacks is usually above 90\%\cite{wang2023experimental}. Even if the text-based CAPTCHAs implement some tricks like interference lines, distortion, overlapping, rotation, or complex backgrounds to resist the segmentation and recognition from attackers, their security remains relatively very low. Adversarial CAPTCHAs employ adversarial perturbations that cannot be perceived by humans, thus deceiving convolutional neural networks and reducing the success rate of attackers\cite{ye2020using,osadchy2017no,shao2022robust}. However, even with the adversarial CAPTCHAs, the security cannot be guaranteed when the techniques such as perturbation removal and adversarial training\cite{bai2021recent,shafahi2019adversarial} are adopted.

Image-based CAPTCHA schemes typically require users to select one or more images from a set of candidates. Common image-based CAPTCHA schemes will superimpose distorted and rotated characters on complex background images. Image-based CAPTCHAs, with larger image sizes and more complex content characteristics, achieve higher security compared with pure text-based CAPTCHAs and have been widely used on the Internet. However, under attack algorithms based on segmentation\cite{wang2021make} or Nonsegmentation \cite{nian2022deep}, the probability of successful attacks can still reach over 70\%. To this end, we need to employ some new techniques to enhance the security of the CAPTCHAs.

This paper proposes a novel image-based CAPTCHAs generation scheme called Diff-CAPTCHA, based on denoising diffusion models\cite{ho2020denoising}. Different from most other existing image-based CAPTCHAs, Diff-CAPTCHA does not simply overlay some text onto the background images. Instead, it treats the background images and characters of the CAPTCHAs as a whole to guide the generation process of the diffusion model. This mechanism weakens the edge features of characters that are usable for machine learning, enhances the diversity of character features in the CAPTCHA, deeply integrates the CAPTCHA characters and background images, and increases the difficulty of attack algorithms.

The main contributions of this paper are summarized as follow:

1. An image-based CAPTCHA generation scheme based on diffusion models is proposed. This scheme uses the  denoising diffusion models to redraw the text in CAPTCHA images, enhancing the diversity and structural complexity of the CAPTCHA characters. The reverse process of the diffusion model is optimized to improve the image generation quality in CAPTCHA application scenarios and achieve more stable generation results to ensure the usability of CAPTCHA.

2. The security and usability of Diff-CAPTCHA are comprehensively validated. Object detection algorithms, two-step attacks, and other mainstream CAPTCHA recognition methods are used to simulate the attacks on Diff-CAPTCHA which is compared with several image-based CAPTCHAs, demonstrating the security of Diff-CAPTCHA. In addition, the usability of the CAPTCHA scheme is validated through manual testing.

The rest of this paper is organized as follows: Section 2 introduces the related works; Section 3 introduces the detailed architecture of our proposed Diff-CAPTCHA scheme; Section 4 conducts extensive experiments to validate the security and usability of Diff-CAPTCHA. Finally, Section 5 concludes this paper.
\section{Related work}

\subsection{CAPTCHA Attack and Defense}
Text-based CAPTCHAs are the most widely used CAPTCHA schemes and are easier to deploy and more stable in performance compared to other alternative CAPTCHA schemes\cite{singh2014survey}. The security mechanisms of traditional text-based CAPTCHAs mainly include anti-segmentation and anti-recognition. Mainstream text-based CAPTCHA schemes include hollow CAPTCHAs, double-layer CAPTCHAs, 3D CAPTCHAs, etc.\cite{chen2017survey}. These CAPTCHA schemes use tricks such as character rotation, blurring, distortion, adding noise interference lines, and character fusion to interfere with machine recognition. At the same time, text-based CAPTCHAs also employ larger character sets, such as Chinese and Japanese, to expand the recognition range of CAPTCHA solvers. In addition, by endowing CAPTCHAs with more diverse character forms, more difficult-to-distinguish objects and backgrounds, the difficulty of CAPTCHA cracking can be increased.

With the continuous development of computer vision technology, CAPTCHA breaking techniques are also getting stronger. In general, CAPTCHA breaking methods can be divided into segmentation-based attacks and end-to-end attacks. In segmentation-based attacks, attackers break CAPTCHAs through multiple steps or stages, typically including preprocessing, segmentation, and recognition. In end-to-end attacks, automatic attack tools take the entire CAPTCHA images as input and directly output the full recognition information of the CAPTCHA\cite{ma2020neural}.

In 2020, Wang et al.\cite{wang2020security} use R-CNN to recognize characters in CAPTCHA images and successfully crack Chinese CAPTCHAs based on large character sets. Wang et al.\cite{wang2021make} propose an effective CAPTCHA solver in 2021, using Cycle-GAN\cite{zhu2017unpaired} for preprocessing of CAPTCHA images and segmenting and recognizing them, achieving over 96\% recognition success rate in various CAPTCHA schemes. In the security validation section, we also develop a two-step attack like\cite{wang2020security}, first using Faster R-CNN\cite{ren2015faster} to segment the CAPTCHA images, and then using the Resnet-50\cite{he2016deep} image classification model to recognize the segmented image blocks.

An end-to-end CAPTCHA cracking algorithm based on attention-based recurrent neural network models is proposed by Zi et al.\cite{zi2019end}, which achieves a high success rate without segmentation or preprocessing. Mocanu et al.\cite{mocanu2022breaking} add capsule networks based on CNN to incorporate spatial information of CAPTCHA features into neural network recognition, effectively addressing the challenges of distortion and character adhesion in CAPTCHA images. In 2021, Nian et al.\cite{nian2022deep} use the Mask R-CNN\cite{he2017mask} object detection algorithm for end-to-end attacks on Chinese CAPTCHA schemes, successfully cracking various text-based CAPTCHAs and image-based CAPTCHAs, achieving a 96\% success rate in a simulated dataset of YiDun-CAPTCHAs. In the security validation section of this paper, a similar end-to-end attack method as\cite{nian2022deep} is used. The difference is that we use the Faster R-CNN\cite{ren2015faster} object detection algorithm because of its faster running speed compared to Mask R-CNN.

Transfer learning-based attack methods\cite{ye2020using,ye2018yet,li2021end} can reduce reliance on training samples. They pre-train a base model on generated samples and then fine-tune the base model using a small amount of labeled data to reduce the complexity and the cost of attacks. Ye et al.\cite{ye2020using,ye2018yet} first develop a CAPTCHA synthesizer to generate simulated CAPTCHAs. They train a base solver on 200,000 images generated by the synthesizer, and then fine-tune the base solver using 200 manually labeled real CAPTCHAs, effectively reducing the cost of attacks and improving accuracy. In 2021, Li et al.\cite{li2021end} use a CAPTCHA synthesizer based on Cycle-GAN to generate synthetic samples for training the base solver, and then use transfer learning techniques to optimize the base recognizer using 500 real CAPTCHAs. The Transfer learning attack in our experiments also is similar to the above methods.

In the face of various segmented-based and end-to-end recognition methods, the security of text-based CAPTCHAs has been severely compromised.

\subsection{Image-based CAPTCHAs}
\begin{figure}[htbp]
	\centering
	\begin{minipage}{0.49\linewidth}
		\centering
		\includegraphics[width=0.8\linewidth]{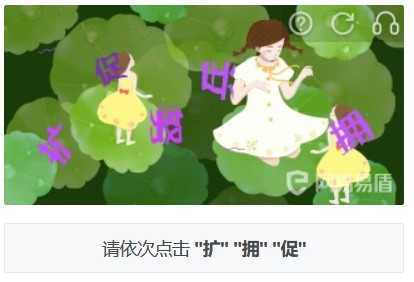}
		\caption{YiDun-CAPTCHA\cite{YIDUNCAPTCHA}}
		\label{YiDun1}%
	\end{minipage}
	\begin{minipage}{0.49\linewidth}
		\centering
		\includegraphics[width=0.8\linewidth]{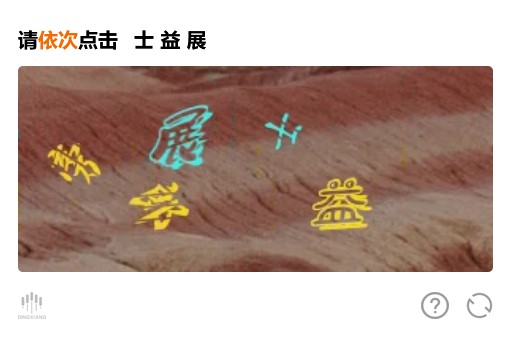}
		\caption{DingXiang-CAPTCHA\cite{DingXiangCAPTCHA}}
		\label{DingXiang1}%
	\end{minipage}
\end{figure}

The image-based CAPTCHAs shown in Figures \ref{YiDun1} and \ref{DingXiang1} adopt several measures to enhance the difficulty of detection and recognition. For example, (a) they use a large character set of Chinese characters, with thousands of characters; (b) they select diverse background images; (c) they use varied fonts and font forms, including hollow and solid changes, shadow changes, color changes, and shape changes like distortion and rotation; (d) they add adversarial perturbations. In addition, related studies have used style transfer neural networks to enhance the security of CAPTCHA images and have achieved good results\cite{cheng2019image,kwon2020captcha}, such as reducing the success rate of breaking the DeCAPTCHA solver to 3.5\% and 3.2\%\cite{kwon2020captcha}.

The CAPTCHAs shown in Figure \ref{YiDun1} and Figure \ref{DingXiang1} essentially involve a simple overlay of foreground characters on background images. The discernible character features remain evident, making them vulnerable to powerful neural network recognition algorithms. For example, Nian et al.\cite{nian2022deep} achieved a 96\% success rate in attacking the simulated NetEase YiDun-CAPTCHA dataset using an end-to-end attack method.

\subsection{Denoising Diffusion Models}
In 2020, Ho et al. propose a denoising diffusion model (DDPM\cite{ho2020denoising}) which is a novel generative neural network with excellent image generation capability. It can generate new images that conform to the distribution of the training set by iteratively denoising the images. The generation process of the diffusion model is diverse, as it can generate images unconditionally from random noise or generate images that meet certain requirements under various guidance. Compared to generative adversarial networks (GANs) or variational autoencoders (VAEs), the diffusion model demonstrates better performance and controllability in guided image generation. Therefore, it quickly becomes one of the mainstream methods in the field of high-resolution image generation\cite{cao2022survey}.

The basic denoising diffusion model consists of a forward process (also known as the diffusion process) and a reverse process. In the forward process, Gaussian noise is gradually added to the image data until the data becomes completely random noise. The reverse process is a denoising process, where noise is gradually removed from the random Gaussian noise by predicting the distribution of the noise. This process continues until an image is generated that approximates the distribution of the real image dataset, like the content used in the training. This achieves the generation of image data by training the neural network to predict the noise in the reverse process, transforming the complex image generation problem into a noise removal problem.

In terms of high-quality image synthesis, DDPM achieves an Inception score of 9.46 and an FID score of 3.17 on the unconditional CIFAR10 dataset. On the 256x256 LSUN dataset, it achieves sample generation quality close to ProgressiveGAN\cite{ho2020denoising}. However, DDPM requires simulating multiple steps of a Markov chain, resulting in high computational costs. To address this issue, Song et al.\cite{song2020denoising} improve DDPM and propose a denoising diffusion implicit model (DDIM), which is a more efficient iterative implicit probabilistic model. DDIM uses a non-Markovian sampling process and is 10 to 50 times faster in sampling compared to DDPM, while maintaining high-quality generated samples. Dhariwal et al.\cite{dhariwal2021diffusion} make a series of improvements to the diffusion model, further improving the sample quality through classifier guidance and controlling gradients to balance the diversity and fidelity of generated samples. Meng et al.\cite{meng2021sdedit} propose the stochastic differential editing method (SDEdit) based on the diffusion model, which uses a stochastic differential equation (SDE) to iteratively denoise and synthesize real images, allowing for a balance between realism and high relevance.

The characteristics of denoising diffusion models make them suitable for image generation tasks in various application scenarios. However, there has been relatively little research on applying diffusion models to CAPTCHA generation. We introduce the denoising diffusion model DDPM to synthesize the background images and characters of the CAPTCHA.

\section{Method}

\subsection{Diff-CAPTCHA}
\begin{figure}[htb]
	\centering
	\includegraphics[scale=0.5]{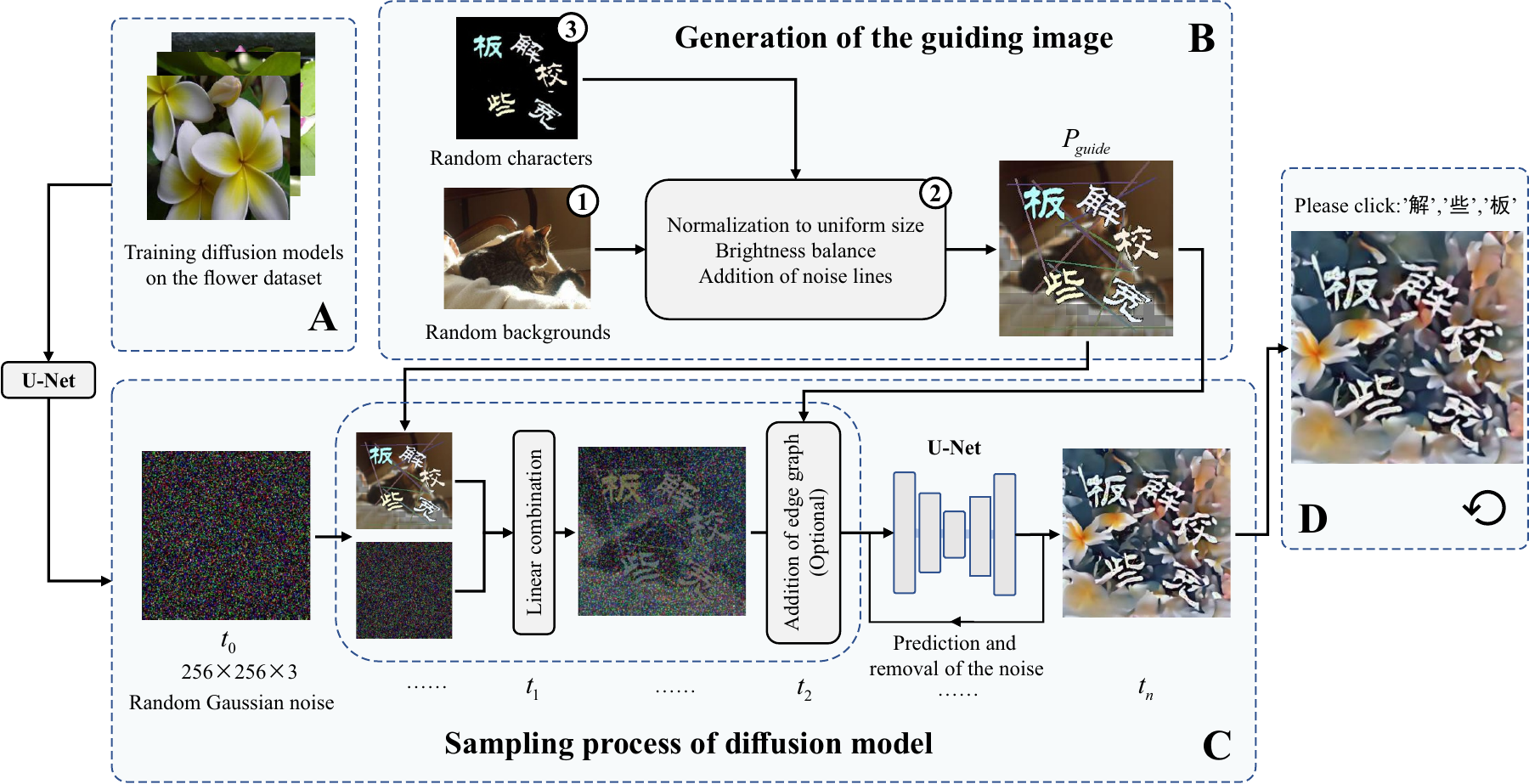}
	\caption{Architecture of Diff-CAPTCHA}
	\label{Architecture of Diff-CAPTCHA}
\end{figure}

The architecture of Diff-CAPTCHA is shown in Figure \ref{Architecture of Diff-CAPTCHA}. As indicated by the dashed box, it consists of the Diffusion Model Training Module (A), Guided Image Generation Module (B), Image CAPTCHA Generation Module (C), and the Application of Image CAPTCHA Deployment (D).

The workflow of Diff-CAPTCHA is as follows: 

First, the Diffusion Model Training Module (A) utilizes a background image dataset, such as the dataset of 724 flower pictures used in the illustration, to train the backbone network U-Net of the diffusion model. Then, during the process of generating the CAPTCHA images, the Guided Image Generation Module (B) first generates the guided images, followed by the Image CAPTCHA Generation Module (C) which generates the final CAPTCHA images.

The following will introduce each module of Diff-CAPTCHA separately.

\subsection{Denoising Diffusion Model Training Module}
\begin{figure}[htb]
	\centering
	\includegraphics[scale=0.5]{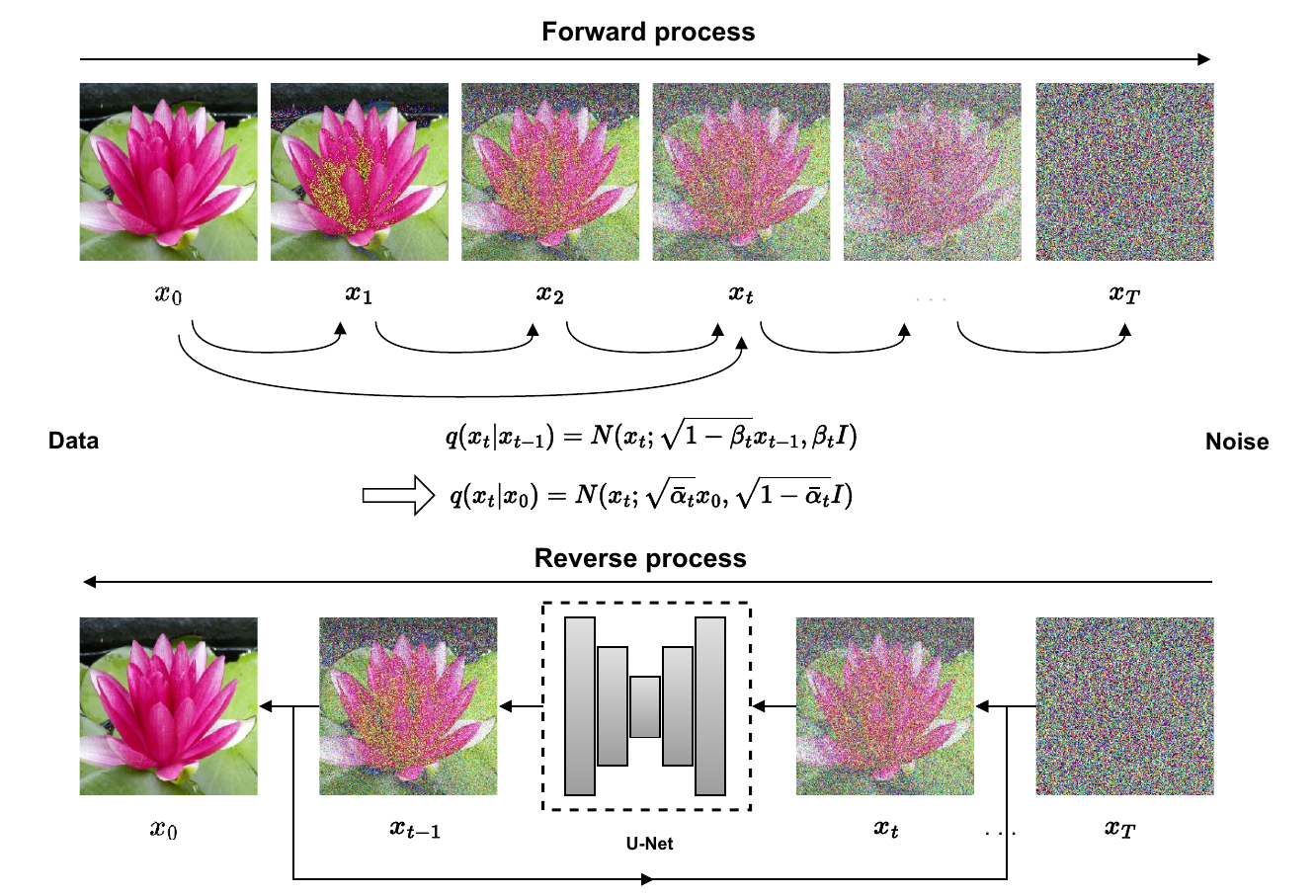}
	\caption{Foward and reverse process of Diffusion Model}
	\label{Foward and reverse}
\end{figure}

As shown in Figure \ref{Foward and reverse}, the diffusion model defines the forward process of gradually adding noise to the original image to generate a noise image that follows a normal distribution, as well as the reverse process of gradually denoising the noise image to restore the original image. Denoising requires using a deep neural network to predict the noise, and in this paper, U-Net is used as the noise prediction network. The purpose of training the diffusion model is to enable the model to predict the previous step's noise information based on the current stage $t$ and the image information $x_t$, which is used in the reverse process of image generation.

\subsubsection{Forward process and reverse process}
The forward process is defined by the denoising diffusion model, which refers to the transformation from the original image to a noisy image. Assuming the original image $x_0$ on the background image dataset follows the distribution $x_0$\textasciitilde $q(x_0)$, the total number of transformations in the forward process is denoted as $T$, and in each round, Gaussian noise following the distribution as shown in Equation (\ref{equ-1}) is added:
\begin{equation}\label{equ-1}
	q(x_t|x_{t-1})=N(x_t;\sqrt{1-\beta_t}x_{t-1},\beta_tI),
\end{equation}
where $x_t$ is the output of the $t$-th round, $x_{t-1}$ is the output of the previous round, and $\beta_t$ is the variance of the $t$-th round, increased with the increase of $t$. It can be obtained by:

\begin{align}\label{equ-2}
\begin{split}
&q(x_t|x_0) = N(x_t;\sqrt{\bar{\alpha_t}}x_0,(1-\bar{\alpha_t})I) \\
\Rightarrow  x_t = &\sqrt{\bar{\alpha_t}}x_0+\sqrt{(1-\bar{\alpha_t})}\epsilon, \qquad where\quad \epsilon \sim N(0,1).
\end{split}
\end{align}

Therefore, $x_t$ can be effectively considered as a linear combination of $x_0$ and random noise $\epsilon$: $x_t=\sqrt{\bar{\alpha_t}}x_0+\sqrt{1-\bar{\alpha_t}}\epsilon$, where the sum of the squares of the coefficients is equal to 1. $\bar{\alpha_t}$ is a coefficient that varies over time.

The reverse process is the step-by-step denoising process, using the deep neural network U-Net to predict the noise changes. The optimization objective of the denoising diffusion model is to make the predicted noise by the deep neural network consistent with the real noise. For more details on the derivation of the forward and reverse processes, please refer to DDPM\cite{ho2020denoising}.

\subsubsection{Training the noise prediction network U-Net}
During the training of the denoising diffusion model, a random training image sample and a random iteration $t$ are selected. A random noise $\epsilon$ is generated according to the standard normal distribution using the trainer. Then, using the forward process Equation (\ref{equ-2}), the noisy image data $x_t$ at the current iteration t is calculated using $x_\theta$ and noise $\epsilon$. xt and $t$ are input into the U-Net neural network to obtain the predicted noise $\epsilon_\theta(x_t, t)$. The L2 loss between the noise $\epsilon$ and predicted noise $\epsilon_\theta$ is calculated and used to update the gradient of the neural network.

The loss function used here is as Equation (\ref{equ-4}):

\begin{equation}\label{equ-4}
L_t=\left \| \epsilon-\epsilon_\theta(\sqrt{\bar{\alpha_t}}x_0+\sqrt{1-\bar{\alpha_t}}\epsilon,t) \right \|^{2}.
\end{equation}

\begin{figure}[htb]
	\centering
	\includegraphics[scale=0.7]{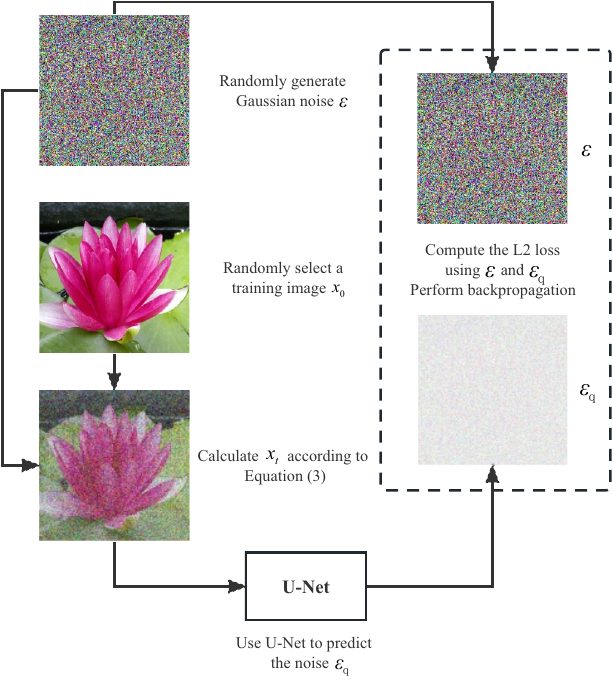}
	\caption{The training process of U-Net}
	\label{Loss of U-net}
\end{figure}

Figure \ref{Loss of U-net} illustrates the training process of the U-Net.

\subsubsection{Architecture of U-Net}

\begin{figure}
	\centering
	\includegraphics[scale=0.5]{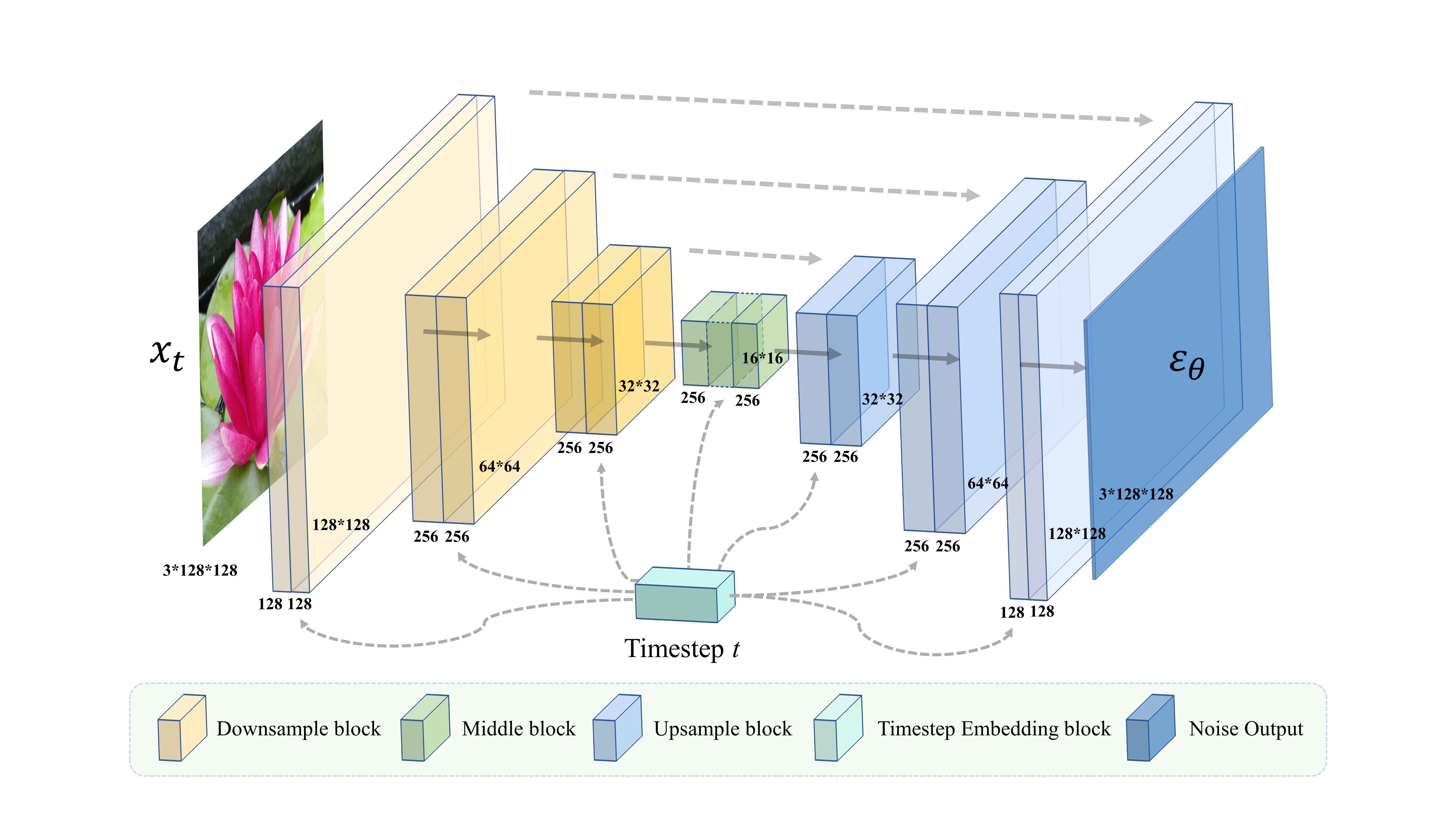}
	\caption{Architecture of U-Net}
	\label{U-net}
\end{figure}

The U-Net\cite{ronneberger2015u} shown in Figure \ref{U-net} is used to predict the noise during the reverse process and reconstruct the images. Since the purpose of using the diffusion model in this paper is to confuse CAPTCHA images and increase the difficulty of automatic attacks, rather than to emphasize the realism of the generated images, an excessively complex model is not necessary. The total number of diffusion steps, Timesteps, is set to 1000, and the current step $t$ is encoded in the network using time embedding to indicate the current step. During model training, the input is a three-channel image $x_t$ with a dimension of 128×128×3. The downsampled convolution kernel (size of 3×3, stride of 2, padding of 1) reduces the edge length of the feature map to half and becomes feature maps of sizes 64×64, 32×32, and 16×16 after three downsampling operations. Correspondingly, three upsampling operations are performed to restore the high-level feature information obtained by the encoder to the size and channel number of the input images. The output of the model is the predicted noise information $\epsilon_\theta(x_t, t)$ at the current step $t$.

To reduce the risk of gradient explosion and gradient vanishing, residual blocks are used in the Downsample block, Middle block, and Upsample block of the model. Self-attention blocks are also added to enhance the model's global modeling capability.
\subsubsection{Datasets for training}
\begin{figure}[htb]
	\centering
	\includegraphics[scale=0.7]{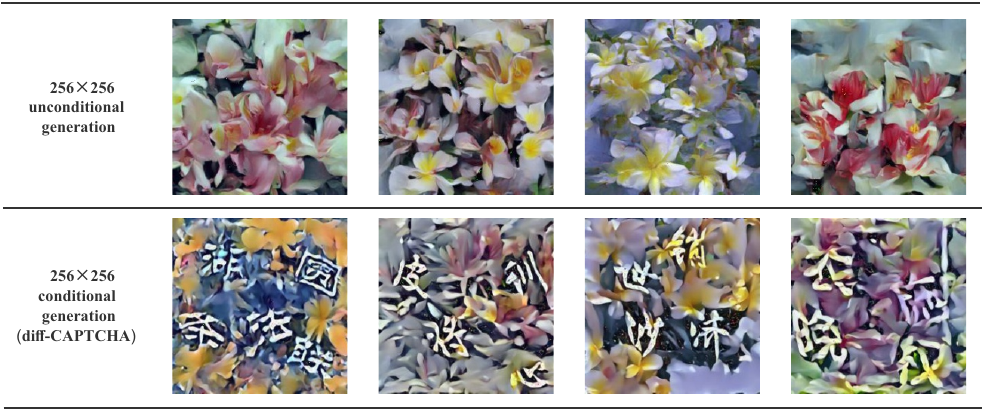}
	\caption{Unconditional generation and conditional generation of Diffusion Model}
	\label{256 generate}
\end{figure}

To ensure the stability of generating CAPTCHA images, only one type of dataset is used during the training of the diffusion model, which includes 724 JPG format images selected from the dataset flower102. The images in the dataset are normalized to a size of 128×128×3 for training the diffusion model. The learning rate during model training ranges from 1e-4 to 2e-4, and the number of training iterations ranges from 200 to 400. The generation process of the Diff-CAPTCHA images are performed on a size of 256×256×3. Figure \ref{256 generate} shows the unconditional and conditionally guided generation results of the diffusion model used in this paper in the size of 256×256×3.

\subsection{Generation of the guiding image}
Hijacking one iteration of the reverse process, adding an extra image to a noisy image by certain coefficients can guide the generation of the final image. This extra image is called the guiding image $P_{guide}$. We utilize this feature of the diffusion models by using an image containing CAPTCHA characters as the guiding image. This ensures that the final CAPTCHA image generated by the diffusion model contains characters, character interference features, background image features, and image features learned by the diffusion model in the training set. These features can be deeply fused, resulting in strong confusion for character detection and recognition.

Therefore, the guiding images include both the CAPTCHA characters and the background interference features. Security enhancement strategies include using a large character set, applying random transformations to the characters, adding random background images, and adding additional interference lines, etc. 

In terms of characters, we use a large-scale Chinese character set, multiple font types, and apply random transformations such as rotation and distortion to the characters. In terms of the background interference of the characters, in addition to adding additional interference lines, we also select 1000 images from the Pascal VOC2012 dataset\cite{everingham2010pascal} as the background. The diverse background images in Pascal VOC2012 dataset can improve the resistance of CAPTCHA images. To facilitate generation, we uniformly convert the background images to a size of 256×256×3.

\begin{figure}[htb]
	\centering
	\includegraphics[scale=0.5]{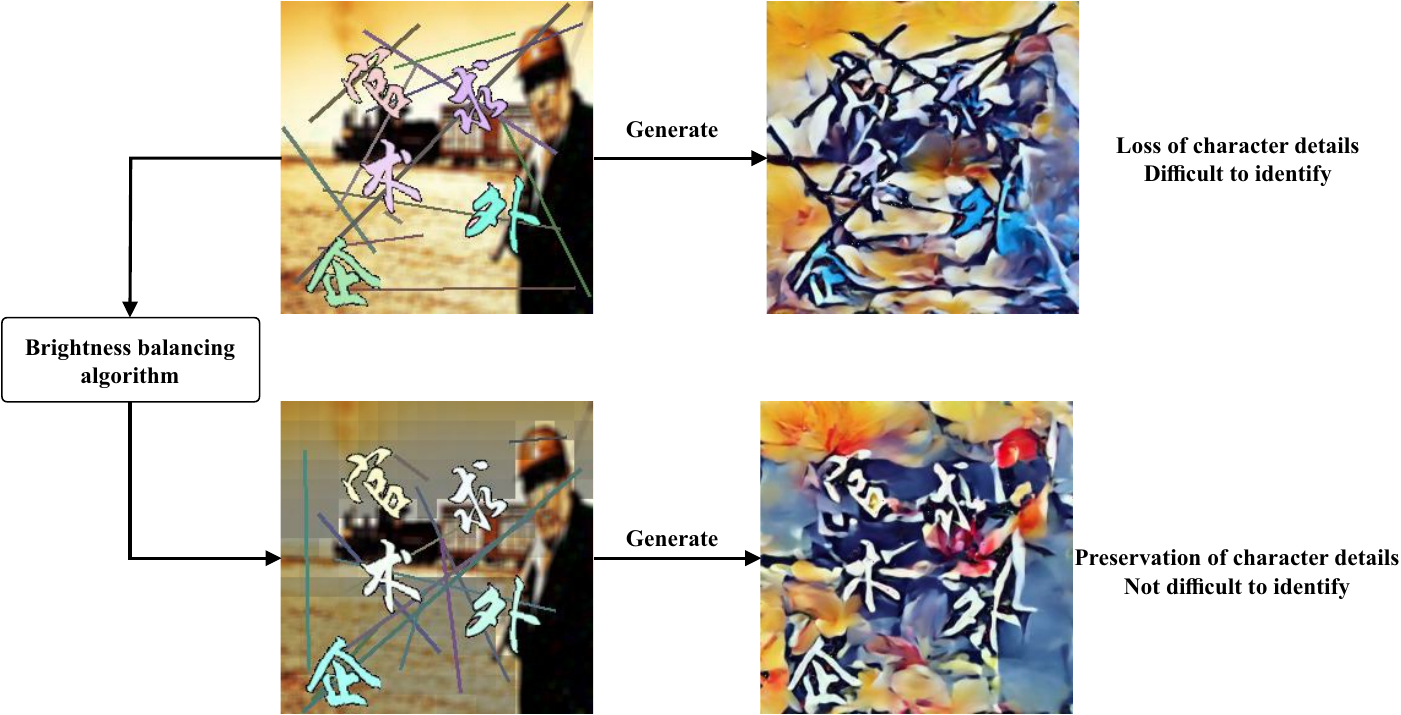}
	\caption{The impact of brightness balancing algorithms on generation results}
	\label{Brightness balance}
\end{figure}

It is worth noting that the color and brightness features of the guiding images have a significant impact on the usability of the final CAPTCHA images. A guiding image with uniform color and brightness can achieve better generation results. However, if the local brightness is too high, it will interfere with the sampling process of the diffusion model, disrupting the character information in the CAPTCHA image and reducing its usability. As shown in Figure \ref{Brightness balance}, the balance of brightness can affect the final generation result. The high contrast in the upper images leads to unclear character features, making them difficult to recognize.

Therefore, we adopt brightness balancing algorithm to reduce the brightness of the high-light areas in images. Specifically, we segment the background of the reference image into small blocks of size n×n and independently calculate the average brightness value $l_{ij}$ in each small block. If this value is higher than the preset threshold $L$, the brightness of that small block is decreased proportionally. As shown in Figure \ref{Brightness balance}, after the brightness balancing algorithm, the originally blurred characters become recognizable by humans.

To sum up, the generation of guided images involves three steps: \textcircled{1}Randomly selecting a real-world image $p_i$ from the background image dataset. \textcircled{2}Preprocessing the image $p_i$, which includes applying Gaussian blur to reduce noise, adjusting the brightness using a brightness balance algorithm for consistent brightness throughout the image, and adding random interfering lines to enhance the CAPTCHA's resistance against attacks. \textcircled{3}Generating the guided image by randomly selecting characters from the character set, applying transformations such as rotation, distortion, and color changes to these characters, and then positioning them randomly into the background image $p_i$ while ensuring minimal overlap between characters. The resulting image is the guided image $P_{guide}$.

\subsection{Generation of Diff-CAPTCHA}
The CAPTCHA generator starts from a randomly generated Gaussian noise image $x_0$ of size 256×256×3 and performs the reverse process of the diffusion model. Then, at a certain iteration stage $t_1$ during the reverse process, the guide image $P_{guide}$ is inserted, and the sampling process continues. The final CAPTCHA image is obtained by denoising using the noise distribution predicted by U-Net. We use the method in DDIM\cite{song2020denoising} to accelerate the sampling process.

Where, the guide image $P_{guide}$ is linearly combined with the noise image $x_{t_1}$ obtained $x_{t_1}^{'}$  after hijacking $t$ rounds. This is done using Equation (\ref{equ-5}).

\begin{equation}\label{equ-5}
	x_{t_1}^{'}=\mu_1 \times x_{t_1}+\mu_2 \times P_{guide}.
\end{equation}

The image $x_{t_1}^{'}$ is then added back to the diffusion model to continue the sampling process. The values of $\mu_1$ and $\mu_2$ will be set differently based on the circumstances. 

Note that in cases where the font is dense, the guide image $P_{guide}$ needs to undergo edge processing using the Canny operator in OpenCV. At the specified timestep $t_2$ (50>$t_2$>$t_1$), the sampling process of the diffusion model is hijacked, and the image $Canny(P_{guide})$ is linearly combined with the image $x_{t_2}$ in the diffusion model. This results in the image 

\begin{equation}\label{equ-6}
	x_{t_2}^{'}=\mu_3 \times x_{t_2}+\mu_4 \times Canny(P_{guide}).
\end{equation}

The image $x_{t_2}^{'}$ is then added back to the sampling process until the predetermined $T$ rounds are completed ($T$ is set to 50). $\mu_3$ and $\mu_4$ are constants. The edge information obtained by the Canny operator can effectively retain character information and maintain the usability of the CAPTCHA.

The image output by the diffusion model is the final CAPTCHA image.

The parameters $t_1$,$t_2$,$\mu_1$,$\mu_2$,$\mu_3$,$\mu_4$, etc. in the generation process above will affect the generation effect of the image. It requires experimental tuning to strike a balance between security and usability.
\section{Evaluation of Security and Usability}

\subsection{Experimental setup}
The training, inference for the image-based CAPTCHAs generation model and the CAPTCHA attack model, are conducted on a workstation with an RTX4090 graphics card, an i7-12700 CPU, 64G of memory, and Windows10 operating system. The programming language used is Python 3.9, and the PyTorch version is 11.7.
\subsection{Evaluation Indicators}
The CAPTCHA breaking program in this paper is based on object detection models and image classification models. To evaluate its attack performance, indicators such as mean Average Precision($mAP$), Attack Success Rate($ASR$), Single Char Attack Success Rate($SCASR$), and Mean Attack Speed($MAS$) are used to measure its performance.

Mean Average Precision ($mAP$) is the average of the average precisions($AP$) for each class. $AP$ is the area under the Precision-Recall(P-R) curve, where Precision is the accuracy and Recall is the recall. The calculations are as follows:

\begin{equation}\label{equ-7}
	Precision=\frac{TP}{TP+FP},
\end{equation}
\begin{equation}\label{equ-8}
	Recall=\frac{TP}{TP+FN},
\end{equation}
where $TP$(True Positive) is the number of correctly predicted positive samples with an $IoU$ greater than 0.5. $FP$(False Positive) is the number of falsely predicted positive samples or predicted positive samples with an $IoU$ less than 0.5. $FN$(False Negative) is the number of objects that were not detected. $IoU$(Intersection-over-Union) represents the ratio of the intersection area to the union area between the predicted region and the ground truth region. A higher $IoU$ indicates a closer match between the predicted region and the ground truth region.

The range of $mAP$ is between 0 and 1, with a larger value indicating a stronger model performance. However, it is important to note that although $mAP$ can evaluate the overall performance of an object detection model, it may not accurately represent the true attack capability of the model in practical CAPTCHA attack scenarios.

Attack Success Rate($ASR$) is defined as the rate at which the breaking program successfully finds all the characters required by the CAPTCHA. In the case of image CAPTCHAs with hint information, an attack is considered successful if the detection box with a confidence higher than 40\% contains all the characters required by the hints and has an $IoU$ greater than 0.5. The calculation of $ASR$ is as follows:

\begin{equation}\label{equ-9}
	ASR=\frac{N_s}{N},
\end{equation}
where $N_s$ is the number of successful attacks, and $N$ is the total number of CAPTCHAs tested. In this paper, each attack experiment uses 1000 independently generated CAPTCHA images, so $N$ is a fixed value of 1000.

Single Char Attack Success Rate($SCASR$) evaluates the classification accuracy of the recognition model on individual character images after segmentation in a two-step attack. It intuitively reflects the classification ability of the image classification model. The calculation of $SCASR$ is as follows:

\begin{equation}\label{equ-10}
	SCASR=\frac{N_{sc}}{N_{allc}},
\end{equation}
where $N_{sc}$ is the total number of successfully recognized characters in all CAPTCHA images, and $N_{allc}$ is the total number of characters in all CAPTCHA images.

Mean Attack Speed($MAS$) measures the time consumed by the attack model to perform one attack. The time consumption of the attack model reflects its practicality. In this paper, the average time used for each attack in the testing phase will be recorded, in seconds.

\subsection{Security Evaluation}
To evaluate the security of Diff-CAPTCHA, we implement end-to-end and two-step CAPTCHA attack methods, and compare Diff-CAPTCHA with multiple baseline CAPTCHAs.

\subsubsection{Diff-CAPTCHA and baseline CAPTCHAs}
Firstly, we introduce datasets of Diff-CAPTCHA and baseline CAPTCHAs, including a small-scale commercial YiDun-CAPTCHA dataset, a large-scale simulated YiDun-CAPTCHA dataset, and style transfer CAPTCHA datasets based on CycleGAN.

\textbf{Diff-CAPTCHA Datasets}

We use the Diff-CAPTCHA scheme proposed in Section 3 to automatically generate a dataset of 10,000 CAPTCHA images, and record all structure information of the CAPTCHA characters including the position and the type of the characters. 80\% of the images are selected as the training set and 20\% as the test set. Additionally, 1,000 images are generated for subsequent security testing. 

\begin{figure}[htb]
\centering
\includegraphics[scale=0.8]{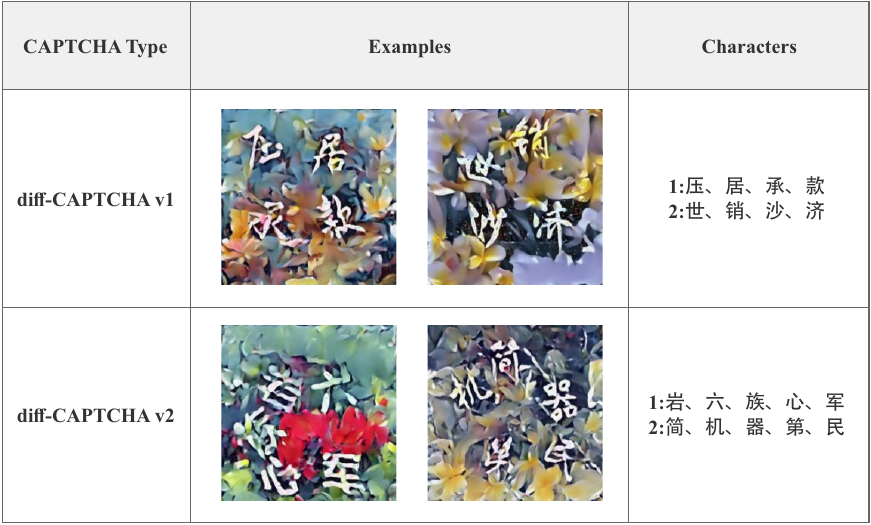}
\caption{Examples of Diff-CAPTCHA}
\label{Examples of Diff}
\end{figure}

Diff-CAPTCHA is designed in v1 and v2 versions. The v1 version has 46,378 labeled Chinese characters, and the v2 version has 56,886 labeled Chinese characters. The v2 version introduces more randomness to enhance security. Figure \ref{Examples of Diff} shows examples of Diff-CAPTCHA images.

\textbf{Baseline CAPTCHA Datasets}

Due to the high cost of collecting and labeling Image-based CAPTCHAs on the Internet, this study uses three programmatically generated CAPTCHA datasets as baseline CAPTCHAs, including simulated YiDun-CAPTCHA and two Image-based CAPTCHAs based on style transfer. The format and partition of each baseline dataset is the same as the Diff-CAPTCHA dataset. Additionally, a small-scale commercial YiDun-CAPTCHA is collected and labeled for attack testing using transfer learning.

The baseline CAPTCHAs used in this paper are as follows:
\begin{itemize}
	\item High imitation of YiDun-CAPTCHA. NetEase YiDun is a well-known CAPTCHA service provider in China. In this paper, a high imitation of YiDun-CAPTCHA is generated with similar visual effects in terms of random deformation of fonts.
	\item Security-enhanced CAPTCHA based on CycleGAN. Style transfer algorithms, including CycleGAN, can change the image style while retaining the original meaning of characters in CAPTCHA images, which can improve the security of CAPTCHA images\cite{zhu2017unpaired}. We refers to the methods proposed by\cite{cheng2019image} and\cite{kwon2020captcha} to generate two versions of security-enhanced CAPTCHA, namely v1 and v2. The character size of the v2 version is random. Please note that these CAPTCHA schemes are generated by ourselves and can not reflect the actual performance of the schemes described in papers above. 
\end{itemize}

\begin{figure}[htbp]
	\centering
	\includegraphics[scale=0.6]{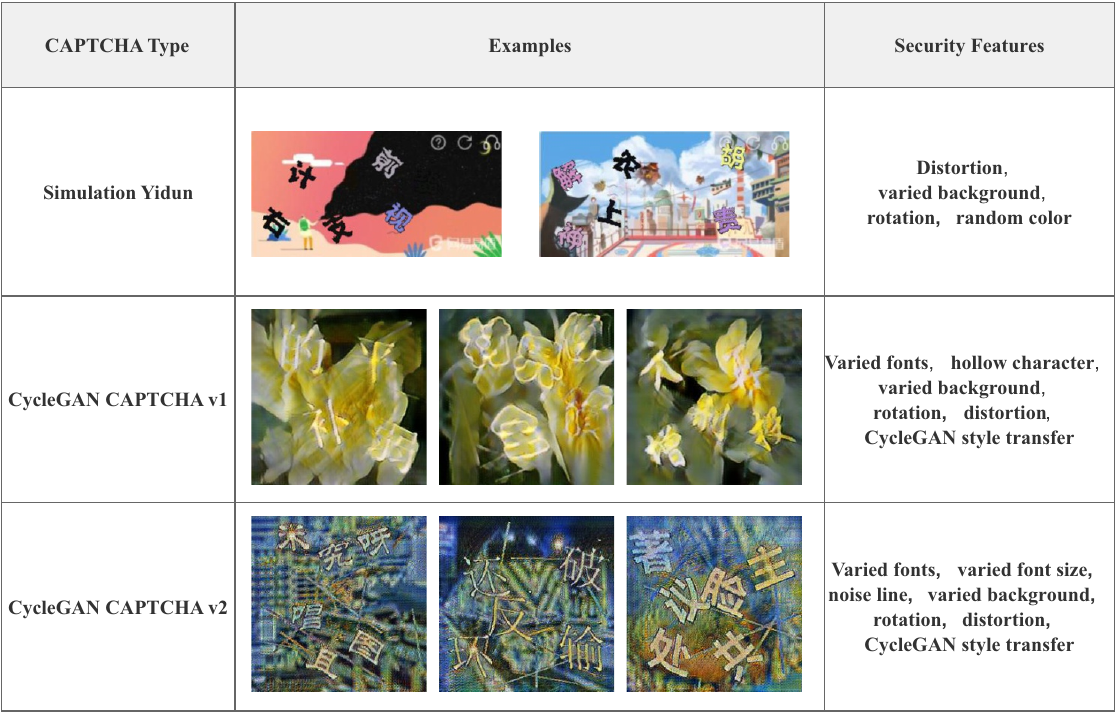}
	\caption{Examples of baseline CAPTCHAs}
	\label{Examples of other sets}
\end{figure}

Figure \ref{Examples of other sets} shows examples of CAPTCHA images from the three baseline CAPTCHAs. Table \ref{Parameters of Different Datasets} lists the parameters of the experimental datasets in this paper.

\begin{table}[!htbp]
	\centering
	\caption{Parameters of Different Datasets}
	\label{Parameters of Different Datasets}
\begin{tabular}{cccccc}
	\toprule 
	Dataset & Sum & Size & Font number & Font size & Length \\ 
	\midrule 
	Diff-CAPTCHA v1 & 10000 & 256×256×3 & 5 & 75 & 4\textasciitilde6\\
	Diff-CAPTCHA v2 & 10000 & 256×256×3 & 4 & 55\textasciitilde70 & 4\textasciitilde6\\
	Simulation YiDun-CAPTCHA & 10000 & 400×200×3 & 2 & 38\textasciitilde45 & 4\textasciitilde6\\
	CycleGAN CAPTCHA v1 & 10000	& 256×256×3	& 5	& 75 & 4\textasciitilde6\\
	CycleGAN CAPTCHA v2 & 10000 & 256×256×3	& 4	& 55\textasciitilde70 & 4\textasciitilde6\\
	\bottomrule 
\end{tabular}
\end{table}

\subsubsection{End-to-end object Detection Attack Test}
The Faster R-CNN-FPN model is selected to crack CAPTCHAs in an end-to-end manner. This model is a widely used object detection algorithm. It incorporates the Feature Pyramid Network(FPN)\cite{lin2017feature} into the Faster R-CNN\cite{ren2015faster} architecture and uses ResNet-50 as the backbone network. It significantly improves the detection accuracy and the ability to detect multi-scale objects compared to the early version of Faster R-CNN, and improves the detection speed. 

In the attack test here, Faster R-CNN-FPN treats each character as a target. There are a total of 1001 kinds of targets, including 1000 Chinese characters from the CAPTCHA character set and the background. In the inference state, the input of Faster R-CNN-FPN is the CAPTCHA image, and the output is information about the detected character, including type, position coordinates, and confidence. In real attack scenarios, the cost of data annotation is usually high. Therefore, we also consider the attack performance under different dataset sizes, setting up three groups of datasets containing 5000, 7500, and 10000 CAPTCHA images respectively. The numbers of images used to train the attack model are 4000, 6000, and 8000 for each group. The total number of epochs is 20, the initial learning rate is 0.005, the learning rate is reduced every 5 epochs, and the batch size is 8. 

During the attack, the CAPTCHA images are first inputted into Faster R-CNN-FPN, which returns information about all detected boxes in the images, including coordinates, object types, and confidence. Then, the detected boxes are compared with the ground truth boxes. If the detected objects contain all the characters indicated in the hints and the $IoU$ is greater than 0.5, the attack is considered successful. 

\begin{figure}[htbp]
	\centering
	\includegraphics[scale=0.65]{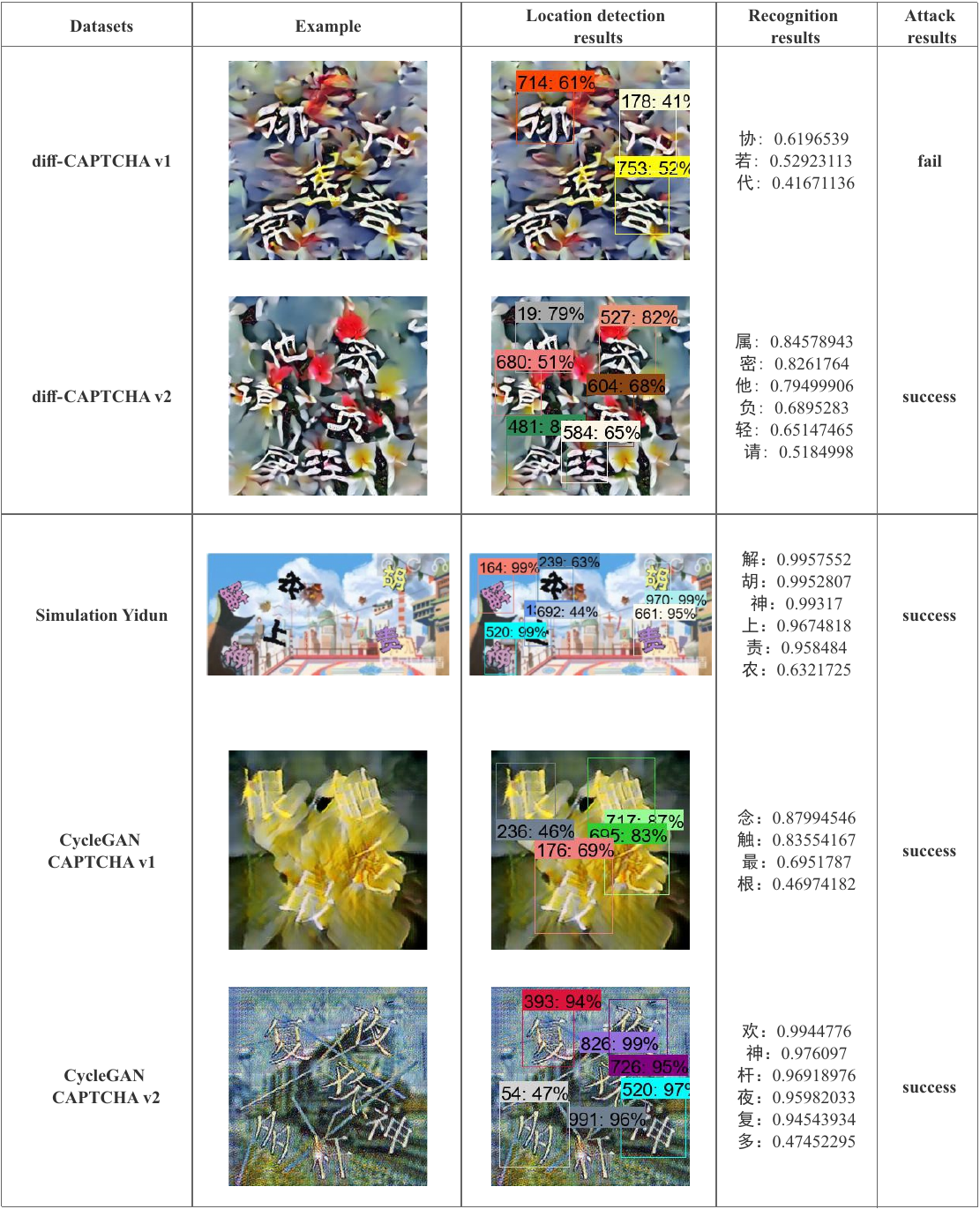}
	\caption{End-to-end object detection attack method}
	\label{End-to-end attack}
\end{figure}

Figure \ref{End-to-end attack} illustrates the attack process, with the index numbers of the detection boxes representing the indexes of the characters, and the confidence scores displayed on the right side.

\begin{table}[!htbp]
	\centering
	\caption{End-to-end object detection attack results}
	\label{End-to-end object detection attack results}
	\begin{tabular}{cccc}
		\toprule 
		Datasets & $mAP$ & $ASR/\%$ & $MAS/s$ \\ 
		\midrule 
		Diff-CAPTCHA v1 & 0.479 & 15.6 & 0.0472 \\
		Diff-CAPTCHA v2 & 0.458 & 6.7 & 0.0536 \\
		CycleGAN CAPTCHA v1 & 0.914 & 79.2 & 0.0516 \\
		CycleGAN CAPTCHA v2 & 0.966	& 84.0	& 0.0501 \\
		Simulation YiDun-CAPTCHA & 0.990 & 86.5	& 0.0474 \\
		\bottomrule 
	\end{tabular}
\end{table}

\begin{figure}[htbp]
	\centering
	\includegraphics[scale=0.45]{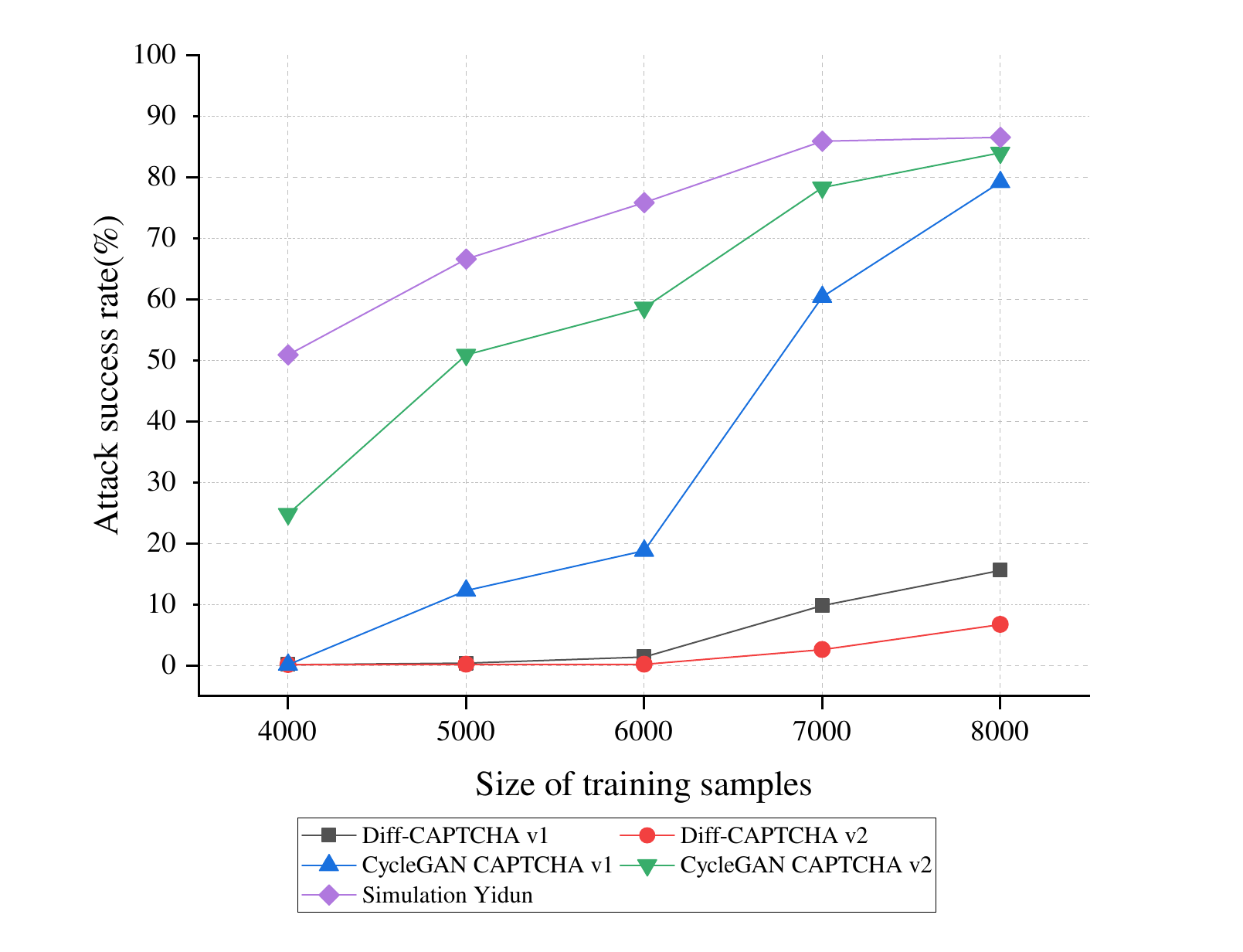}
	\caption{End-to-end attack success rates under different sample sizes}
	\label{End-to-end figure}
\end{figure}

Table \ref{End-to-end object detection attack results} and Figure \ref{End-to-end figure} record the $mAP$, average attack rate, and average attack speed of the Faster R-CNN model on different CAPTCHAs. On smaller datasets, Faster R-CNN can’t effectively attack the five CAPTCHAs. However, given larger training datasets, the success rate of the attacks has increased. The average attack rates for attacking Diff-CAPTCHA are still low, while the average attack rates for other CAPTCHAs exceed 70\%. Therefore, compared to the baseline CAPTCHAs, Diff-CAPTCHA has stronger resistance against end-to-end attacks.
\subsubsection{Transfer Learning-based Attack Test}

For CAPTCHAs in real commercial scenarios, training end-to-end attack models requires a large number of accurately labeled CAPTCHA images. For example, the attack method described in Section 4.3.2 requires a dataset of nearly ten thousand accurately labeled CAPTCHAs, which is often unacceptable due to its high labor cost. Therefore, we used a transfer learning-based attack method when comparing Diff-CAPTCHA and YiDun-CAPTCHA.

The transfer learning attack first pretrains the attack model using a high-quality imitation dataset and then fine-tunes the model using a set of real-world labeled CAPTCHA images to enhance the attack performance. Our transfer learning attack method still uses Faster R-CNN as the detection tool.

\begin{figure}[htbp]
	\centering
	\includegraphics[scale=0.45]{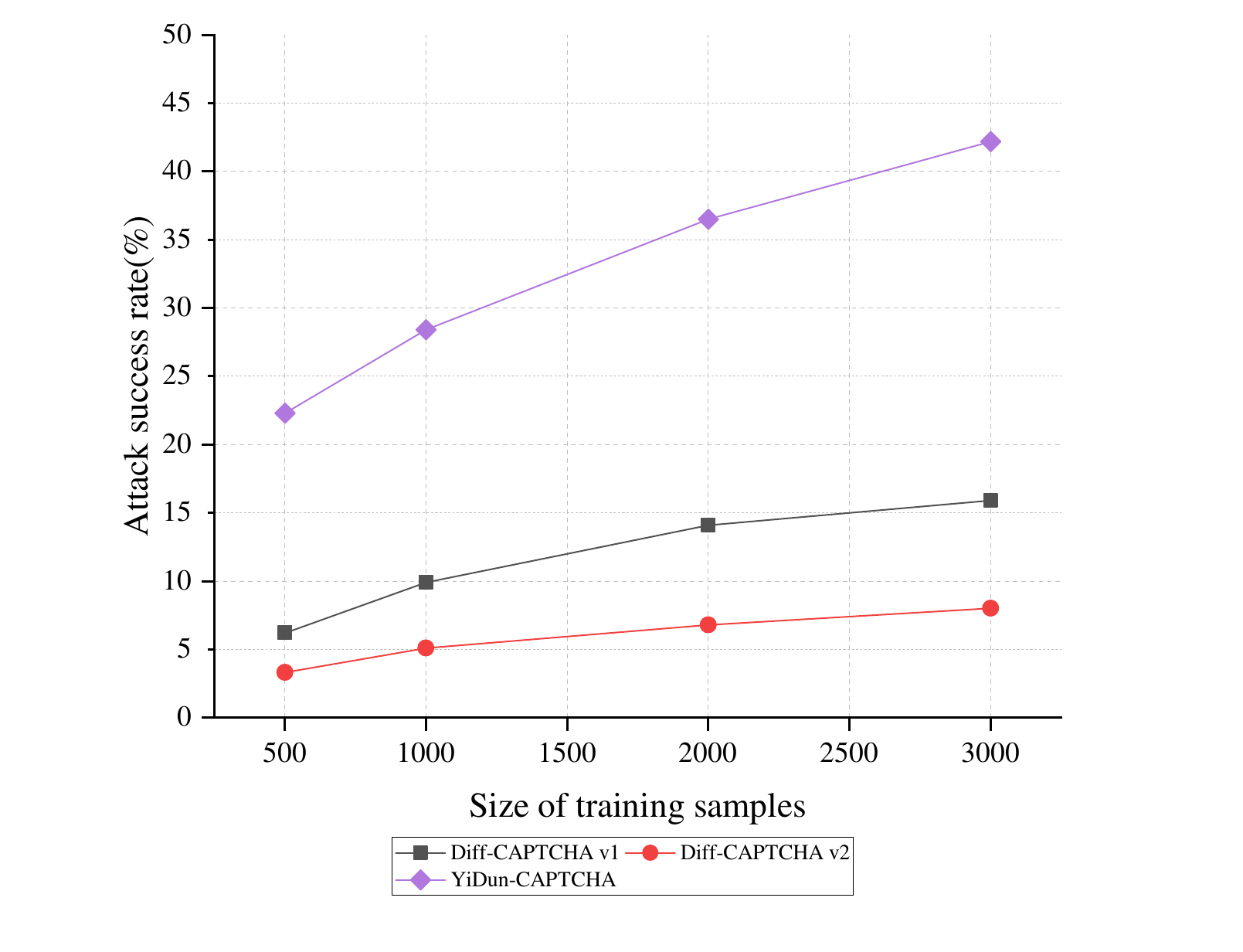}
	\caption{Transfer Learning-based Attack success rates under different sample sizes}
	\label{Trans attack figure}
\end{figure}

For the actual commercial YiDun-CAPTCHA, we first collect 32,000 CAPTCHA images and their corresponding prompts, and compile a character set of 1,874 Chinese characters that appeared. Then, by collecting background images from YiDun-CAPTCHA and simulating its security mechanisms, we generate 20,000 highly similar CAPTCHAs, which are used as the training set for pre-training the Faster R-CNN model. The pre-training is conducted for 20 epochs with an initial learning rate of 0.005, and the learning rate is reduced every 5 epochs. The batch size is set to 8. The trained model achieves a mean Average Precision ($mAP$) of 0.908 on the validation set, demonstrating high accuracy.

Subsequently, a manually annotated dataset of 3,000 real-world YiDun-CAPTCHA images is used to fine-tune the Faster R-CNN for 10 epochs. The fine-tuned model is then subjected to an attack test on 1,000 real-world YiDun-CAPTCHA images. The success rate of attacks significantly improved compared to before, with an overall attack success rate reaching 42.2\%.

As a comparison, the same approach is used for transfer learning attacks on Diff-CAPTCHA. A total of 20,000 images are generated with the same font as Diff-CAPTCHA but without using the diffusion model, and these images are used as the training set for pre-training the Faster R-CNN model. Subsequently, the model is fine-tuned using 500, 1,000, 2,000 or 3,000 samples. The experimental results show that the attack success rate of Diff-CAPTCHA v1 increase to 15.9\%, and the success rate of v2 increase to 8.0\%. The test results are in Figure \ref{Trans attack figure}.

The overall attack success rate of Diff-CAPTCHA is relatively low, indicating that the security of Diff-CAPTCHA is better than the actual commercial YiDun-CAPTCHA under the same testing conditions.

\subsubsection{Two-step Attack Test}
\begin{figure}[htbp]
	\centering
	\includegraphics[scale=0.65]{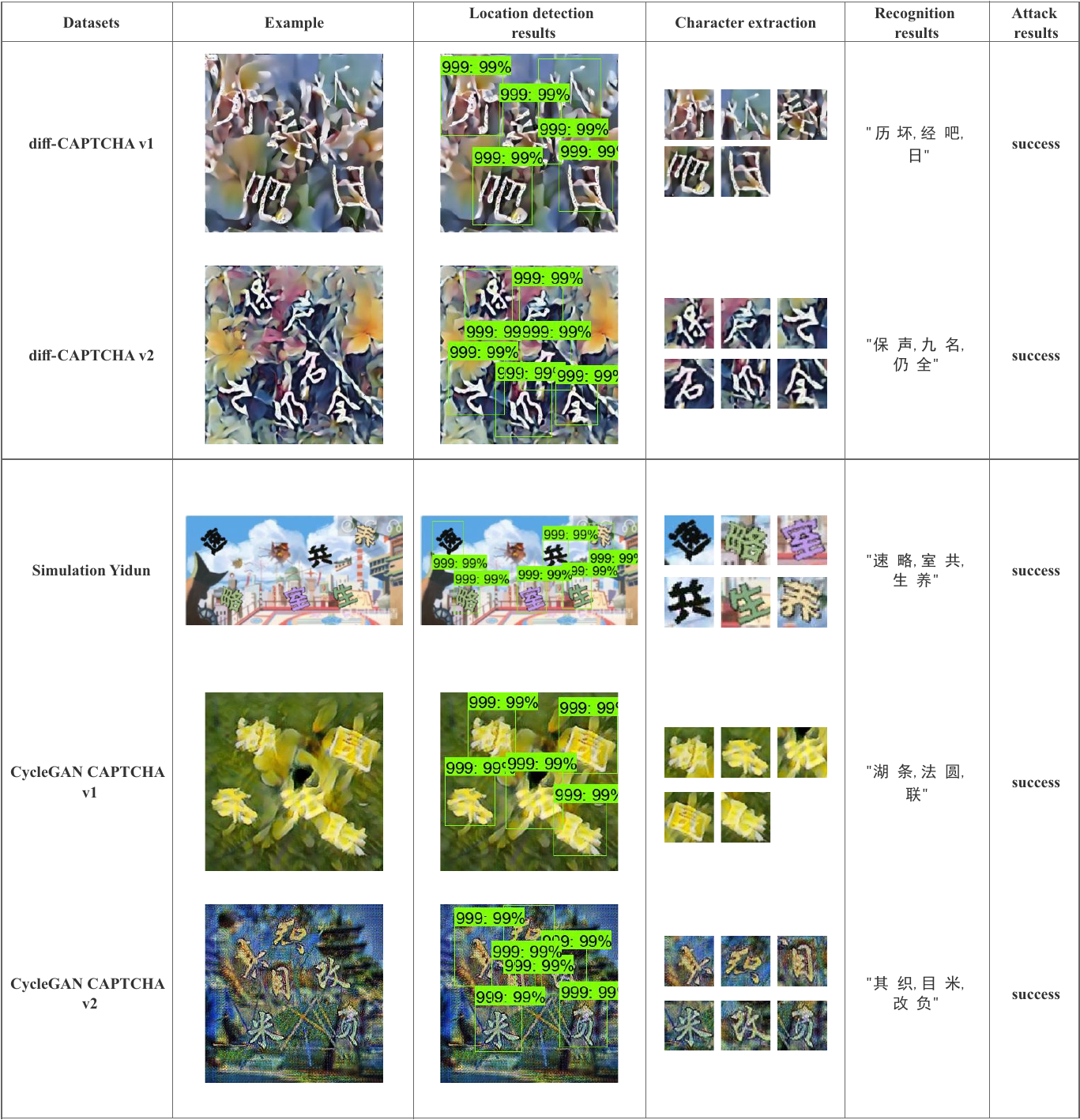}
	\caption{Two-step attack method}
	\label{Two-step attack figure}
\end{figure}

Two-step attack is a commonly used CAPTCHA attack method that includes two steps: segmentation and recognition. Compared to end-to-end attacks, two-step attacks are slightly more complex as they require training two sets of models. In this paper, Faster R-CNN+FPN is used as the segmentation model, and Resnet-50 is used as the recognition model. The attack process is shown in Figure \ref{Two-step attack figure}.

First, Faster R-CNN is trained using datasets consisting of background and objects to segment character blocks from CAPTCHA images. All the characters are grouped into one category. After 10 training iterations, Faster R-CNN achieves a high level of accuracy, with an initial learning rate of 0.005 and a batch size of 8. Then, the characters in CAPTCHA images are segmented based on position information in the datasets to create image classification datasets. ResNet-50 is trained on images normalized to 96×96 pixels, with a learning rate of 0.005 and a batch size of 256.

\begin{figure}[htbp]
	\centering
	\includegraphics[scale=0.45]{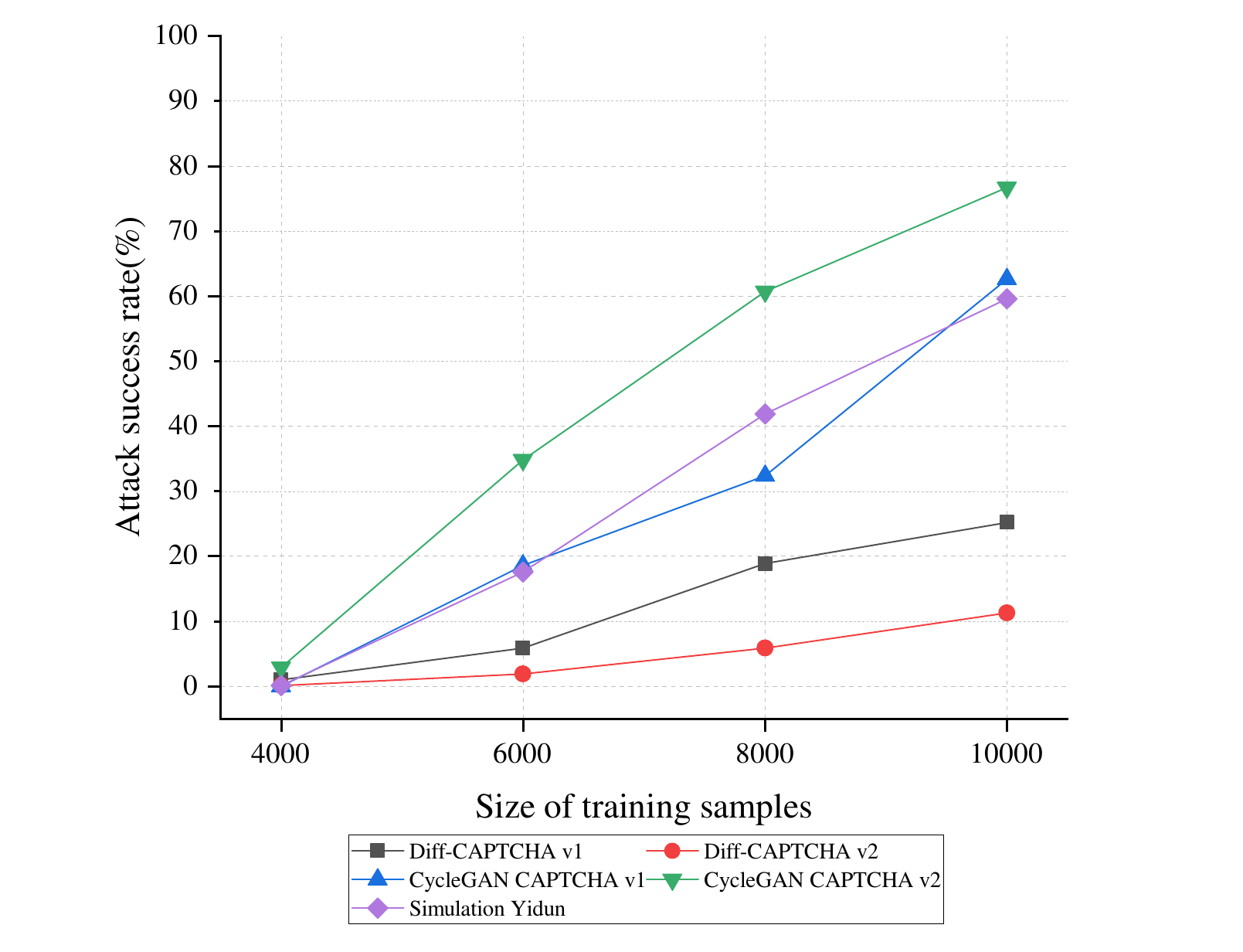}
	\caption{Two-step Attack success rates under different sample sizes}
	\label{Two-step attack ASR}
\end{figure}

\begin{figure}[htbp]
	\centering
	\includegraphics[scale=0.45]{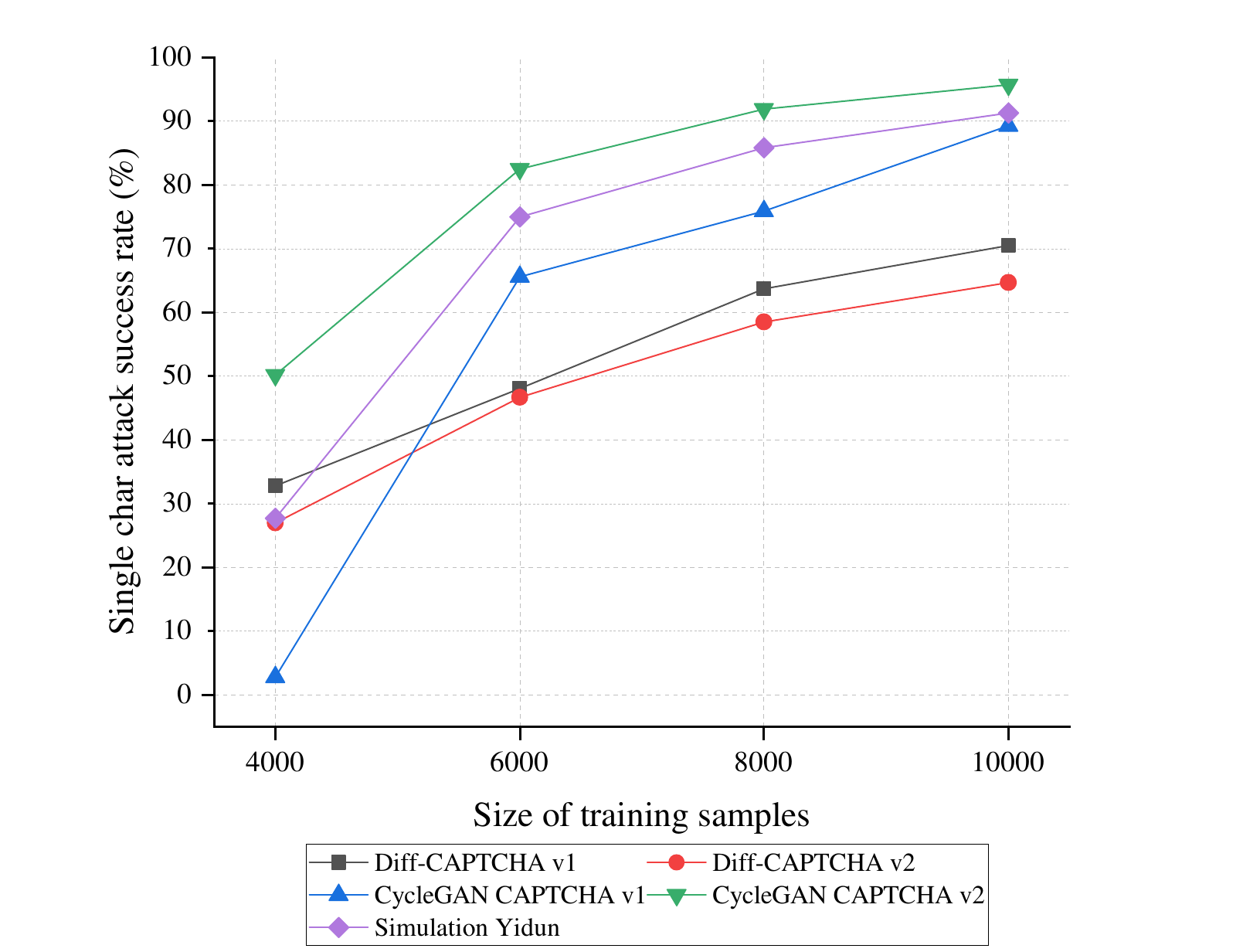}
	\caption{Two-step Single Char Attack success rates under different sample sizes}
	\label{Two-step attack SCASR}
\end{figure}

The attack effects on various CAPTCHAs are shown in Table \ref{Two-step before augmentation}. We later make improvements to the experimental process by applying data augmentation techniques to the training samples when training Resnet-50. We divide the datasets into groups consisting of 4,000, 6,000, 8,000, and 10,000 characters. The attack effects after data augmentation during training are shown in Table \ref{Two-step after augmentation}. Experimental results show that data augmentation during training, including random rotation, cropping, etc., can effectively improve the generalization ability of the Resnet-50 model in CAPTCHA character recognition tasks. It also shows that when attacked by the same attack method, Diff-CAPTCHA has a lower broken rate, indicating that its security mechanism has better security compared to other control schemes. Figure \ref{Two-step attack ASR} and Figure \ref{Two-step attack SCASR} depict line graphs illustrating the test performance.

\begin{table}[!htbp]
	\centering
	\caption{Two-step attack results before data augmentation}
	\label{Two-step before augmentation}
	\begin{tabular}{cccc}
		\toprule 
		Datasets & $SCASR/\%$ & $ASR/\%$ & $MAS/s$ \\ 
		\midrule 
		Diff-CAPTCHA v1 & 45.8 & 6.1 & 0.0668 \\
		Diff-CAPTCHA v2 & 38.6 & 1.1 & 0.0713 \\
		CycleGAN CAPTCHA v1 & 72.0 & 25.3 & 0.0681 \\
		CycleGAN CAPTCHA v2 & 83.2	& 36.4	& 0.0741 \\
		Simulation YiDun-CAPTCHA & 79.7 & 26.2	& 0.0717 \\
		\bottomrule 
	\end{tabular}
\end{table}

\begin{table}[!htbp]
	\centering
	\caption{Two-step attack results after data augmentation. The units of ASR and SCASR are (\%)}
	\label{Two-step after augmentation}
	\begin{tabular}{ccccccccc}
		\toprule
		\multirow{2}[4]{*}{Datasets} & \multicolumn{2}{c}{4000} & \multicolumn{2}{c}{6000} & \multicolumn{2}{c}{8000} & \multicolumn{2}{c}{10000} \\
		\cmidrule{2-9}    \multicolumn{1}{c}{} & \multicolumn{1}{c}{SCASR} & \multicolumn{1}{c}{ASR} & \multicolumn{1}{c}{SCASR} & \multicolumn{1}{c}{ASR} & \multicolumn{1}{c}{SCASR} & \multicolumn{1}{c}{ASR} & \multicolumn{1}{c}{SCASR} & \multicolumn{1}{c}{ASR} \\
		\midrule
		Diff-CAPTCHA v1 & 32.8 & 1.0 & 48.1 & 5.9 & 63.7 & 18.9 & 70.5 & 25.2 \\
		\midrule
		Diff-CAPTCHA v2 & 27.0  & 0.1 & 46.7 & 1.9 & 58.5 & 5.9 & 64.7 & 11.3 \\
		\midrule
		CycleGAN v1 & 2.8  & 0.0    & 65.6  & 18.6  & 75.9  & 32.4  & 89.3  & 62.6 \\
		\midrule
		CycleGAN v2 & 50.2  & 2.9   & 82.5  & 34.8  & 91.9  & 60.7  & 95.7  & 76.7 \\
		\midrule
		Simulation YiDun & 27.7  & 0.1   & 75.0    & 17.6  & 85.8  & 41.9  & 91.3  & 59.6 \\
		\bottomrule
	\end{tabular}%
\end{table}%

\subsection{Usability test}
We develop an interactive image-based CAPTCHA program that requires participants to click on each character in the CAPTCHA images in the correct order according to the prompt text. If all clicks are within the correct character boxes, it is judged as correct. We invited several volunteers to perform usability tests on two CAPTCHA schemes, with 1000 tests for each scheme. The results are shown in Table \ref{Usability test results}.

\begin{table}[!htbp]
	\centering
	\caption{Usability test results}
	\label{Usability test results}
	\begin{tabular}{ccc}
		\toprule 
		CAPTCHA type & Success rate/\% & Average time/s \\ 
		\midrule 
		Diff-CAPTCHA v1 & 93.5 & 9.37 \\
		Diff-CAPTCHA v2 & 90.2 & 11.85 \\
		\bottomrule 
	\end{tabular}
\end{table}

We can seen that the success rate of Diff-CAPTCHA v2 is slightly lower than v1, and the average time consumption is longer than v1, indicating that it is more difficult for humans. The human accuracy rates for Diff-CAPTCHA v1 and v2 are both above 90\%, with an average time consumption of less than 15 seconds, indicating that Diff-CAPTCHA has good usability.

\section{Conclusion and future work}
This paper proposes the Diff-CAPTCHA scheme, which uses a diffusion model to improve the security of the CAPTCHA system. Several attack methods are used to systematically evaluate the security of Diff-CAPTCHA. In the end-to-end attack and two-step attack experiments based on Faster R-CNN, the success rate of Diff-CAPTCHA v2 in end-to-end attack is 6.7\%, and in two-step attack is 11.3\%, lower than other baseline sets. The results show that the diffusion model can be used as a CAPTCHA security mechanism, reducing the success rate of automatic attack in segmentation and recognition, effectively improving the ability of the CAPTCHA system to resist automatic recognition attacks, while maintaining good usability.

In terms of future improvements, the diffusion models use in this paper does not have the ability to generate images guided by labels. Currently, more advanced diffusion models, such as Classifier-Guidance diffusion models\cite{dhariwal2021diffusion} and Classifier-Free models\cite{ho2022classifier}, can generate images that tend to specific targets based on user-provided labels. If this type of diffusion model is used and the system randomly selects one or more labels to guide the generation each time, it can greatly enhance the diversity of generated CAPTCHA images and bring significant improvements in security. However, the training difficulty of the model will also be higher. The combination of diffusion model-based CAPTCHAs and the addition of adversarial examples does not conflict, and the use of multiple security mechanisms can enhance the resistance to attacks. In the future, an improved diffusion model-based CAPTCHA generation system will have better security and faster generation speed. It is hoped that this study can bring new ideas to other CAPTCHA researchers and promote the development of CAPTCHAs.



\bibliographystyle{elsarticle-num-names}
\bibliography{mybib}





\end{document}